\documentclass[acus]{JAC2003}

\usepackage{graphicx}
\usepackage{booktabs}
\usepackage{amssymb}
\usepackage{amstext}

\def\d{{\rm d}}

\begin{document}
\title{TOTEM: PROSPECTS FOR TOTAL CROSS-SECTION AND LUMINOSITY MEASUREMENTS}

\author{M. Deile, CERN, Geneva, Switzerland\\
on behalf of the TOTEM Collaboration}

\maketitle

\begin{abstract}
With the installation of the T1 telescope and the Roman Pot stations at 147\,m
from IP5, the detector apparatus of the TOTEM experiment has been completed 
during the technical stop in winter 2010/2011. After the commissioning of the 
dedicated beam optics with $\beta^{*} = 90\,$m, a first measurement of the 
total pp cross-section $\sigma_{tot}$ and -- simultaneously -- the luminosity
$L$ will be possible in the upcoming running season 2011. The precision
envisaged is 3\% and 4\% for  $\sigma_{tot}$ and $L$, respectively. An 
ultimate beam optics configuration with $\beta^{*} \sim 1$km will later reduce
the uncertainty to the 1\% level.
\end{abstract}

\section{INTRODUCTION}
The motivation and concept of measuring  $\sigma_{tot}$ and $L$ at LHC 
with the method based on the Optical Theorem have been discussed at length
elsewhere~\cite{TDR,JINST}. 
The central equations expressing $\sigma_{tot}$ and $L$
in terms of the measured inelastic and elastic rates, $N_{inel}$ and $N_{el}$, 
and the extrapolation of the differential elastic rate to the optical point, 
$t = 0$, are
\begin{eqnarray}
\sigma_{tot} &=& \frac{16 \pi}{1 + \rho^{2}} \cdot
\frac{dN_{el}/dt |_{t=0}}{N_{el} + N_{inel}}\quad\text{, and} \label{eqn_sigmatot}\\
\mathcal{L} &=& \frac{1 + \rho^{2}}{16 \pi} \cdot 
\frac{(N_{el} + N_{inel})^{2}}{dN_{el}/dt |_{t=0}} \:, \label{eqn_lumi}
\end{eqnarray}
where $\rho = 0.1361\pm0.0015^{+0.0058}_{-0.0025}$~\cite{COMPETE} 
will as a first step be taken from theory. At a later stage, a 
measurement at $\beta^{*} \sim 1$km can be attempted (see last section of this 
article).

This article aims at giving an update on the expected
performance, taking into account recent studies for the reduced centre-of-mass
energy of 7\,TeV and the running experience in 2010.

\begin{figure*}[ht!]
   \centering
   \includegraphics*[width=168mm]{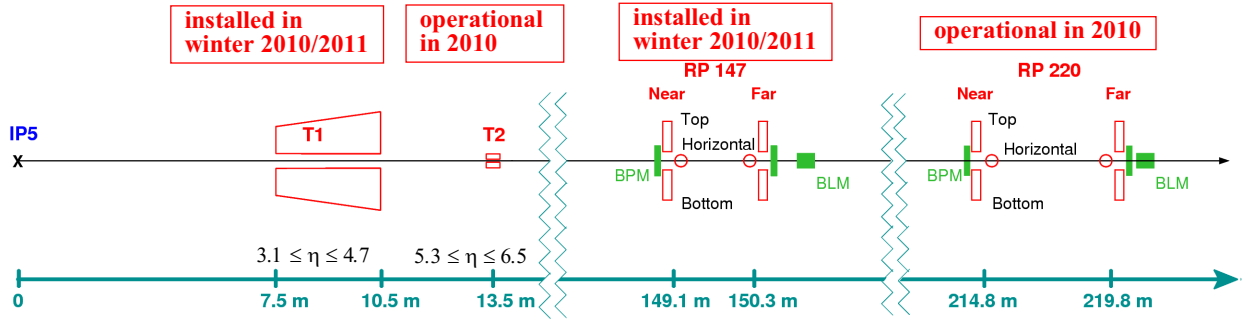}
   \caption{\it Schematic overview of the TOTEM detector configuration. The 
apparatus is symmetric w.r.t. IP5, but only the arm in Sector 5-6 is drawn.}
   \label{fig:apparatus}
\end{figure*}

The Roman Pot (RP) stations at $\pm$220\,m from IP5 and the forward GEM 
telescope T2 had already been fully operational in 2010, whereas the second 
half of the RP spectrometer -- i.e. the $\pm$147\,m stations -- as well as the 
CSC telescope T1 were installed during the technical stop in winter 2010/2011, 
thus completing TOTEM's detector apparatus (Figure~\ref{fig:apparatus}) in 
time for the running season 2011. 

\section{MEASUREMENT OF THE INELASTIC RATE}
The inelastic 
trigger acceptance predicted by simulation for the TOTEM detector configuration
is given in Table~\ref{tab:triggerlosses}.

\begin{table}[h!]
    \setlength\tabcolsep{4pt}
    \caption{\it Trigger losses at $\sqrt{s} = 7\,$TeV, requiring 3 tracks pointing
to the IP. The cross-sections given are the central values of rather wide 
ranges of predictions~\cite{lhcc2010}.}
    \label{tab:triggerlosses}
    \begin{tabular}{|l|c|c|}
        \hline
Event class & $\sigma$ & T1/T1 \\
            &          & trigger and selection loss \\
\hline
Minimum Bias       & 50\,mb & 0.05\,mb \\
Single Diffractive & 12.5\,mb & 4.83\,mb \\ 
Double Diffractive & 7.5\,mb & 1.21\,mb \\
\hline
Total              & 70\,mb & 6.1\,mb \\
        \hline
    \end{tabular}
\end{table}
The trigger losses are dominated by low-mass diffractive events where the 
diffractive system escapes to rapidities beyond the reach of T2, i.e. 
$\eta > 6.5$ or masses $M \lesssim 10\,$GeV.
The missing part of the diffractive mass spectrum can be partially recovered
by extrapolating it according to the empirical relation 
$\frac{dN}{dM^{2}} \propto \frac{1}{M^{n}}$ with $n \approx 2$. 

The systematic uncertainty of this procedure depends on the purity of
the diffractive event sample used for the extrapolation. For example, minimum
bias events misidentified as diffractive events will introduce a bias.
To improve the identification of event topologies, e.g. rapidity gaps, 
CMS detectors at rapidities below 3.1 (central detectors) or above 6.5 
(FSC~\cite{FSC} and ZDC) could be used.

A principal problem of the extrapolation is that 
the $1 / M^{2}$ dependence of the spectrum is not theoretically 
justified~\cite{khoze}. 
There may be sizeable deviations at low masses; even the presence of 
resonances cannot be excluded.
However, an independent handle on Single Diffractive (SD) events can be 
exploited:
At $\beta^{*} = 90\,$m~\cite{lhcc90m}, the leading protons of all diffractive
events with $|t| > 10^{-2}\,$GeV$^{2}$ are detected in the RPs, irrespective 
of the diffractive mass. Thus, SD events whose diffractive system escapes 
detection by T2 will have the signature of {\it unpaired} protons in the 
kinematic region of elastic protons (i.e. $\xi \approx 0$).

After all corrections a total uncertainty of 1\,mb (or 1.4\%) in $N_{inel}$
is expected.

\section{MEASUREMENT OF THE ELASTIC RATE AND EXTRAPOLATION TO t = 0}
\begin{figure*}[ht!]
   \centering
   \includegraphics*[width=145mm]{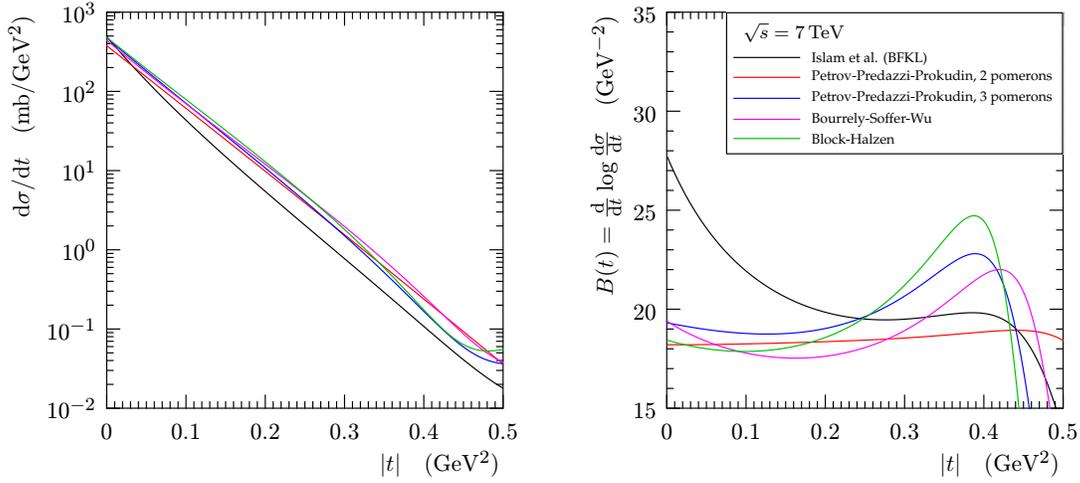}
   \caption{\it Left: elastic differential cross-section at low $|t|$ as 
predicted by different models~\cite{elastic}. Right: exponential slope of (left).}
   \label{fig:sigmaelastic}
\end{figure*}

Predictions~\cite{elastic} for the differential cross-section of elastic pp 
scattering at $\sqrt{s} = 7\,$TeV in the low-$|t|$ region
are displayed in Figure~\ref{fig:sigmaelastic}(left).  
Deviations from the apparent exponential behaviour become visible when the 
exponential slope, $B(t) = \frac{\d}{\d t} \ln \frac{\d \sigma}{\d t}$ is 
plotted (Figure~\ref{fig:sigmaelastic}, right). This non-constant $B(t)$ has
to be taken into account in the extrapolation fit for obtaining
$\frac{\d \sigma}{\d t} (t = 0)$. The procedure is identical to the one 
described in~\cite{JINST} for $\sqrt{s} = 14\,$TeV. 

\begin{figure}[h!]
   \centering
   \includegraphics*[width=6.5cm,angle=270]{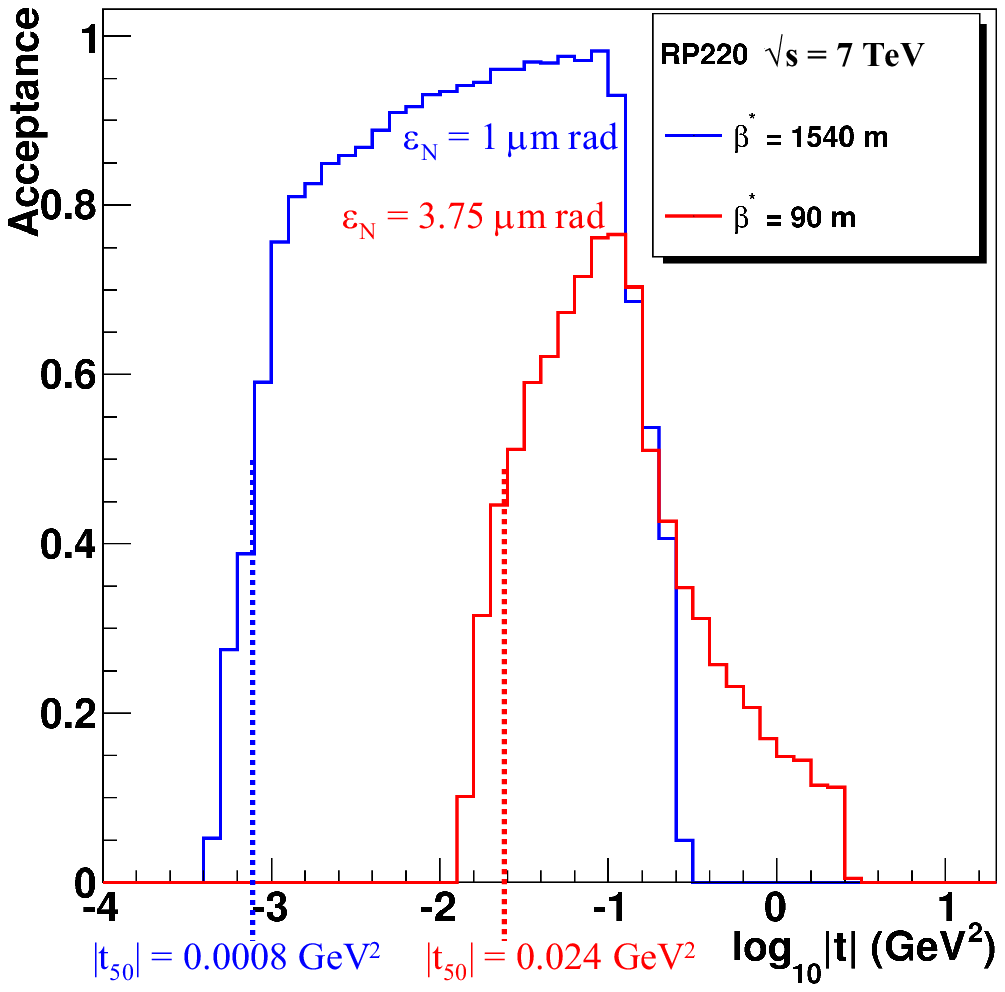}
   \caption{\it Acceptance of the RP220 station for elastically scattered protons
in terms of $t$ for the two high-$\beta^{*}$ optics and for a detector-beam 
distance of $10\,\sigma$. The point where the 
acceptance reaches 50\% on the lower-$|t|$ side is called $t_{50}$; it 
characterises the typical minimum $|t|$ useable for extrapolation purposes.}
   \label{fig:elasticaccept}
\end{figure}

\begin{figure}[h!]
   \centering
   \includegraphics*[width=7cm]{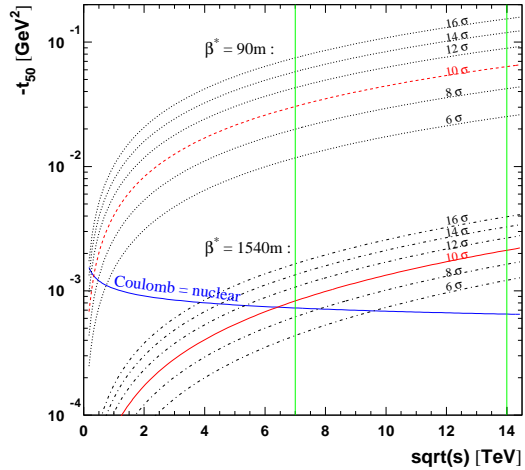}
   \caption{\it $|t|$-value at which the RP220 acceptance reaches 50\%, as a 
function of the centre-of-mass energy, for different distances between the 
RP window and the beam. The sensitive detector volume begins $\delta=0.5$\,mm further away
(due to window thickness and window-detector gap). The $|t_{50}|$-values take 
$\delta$ into account.
The upper and lower blocks of curves represent the $\beta^{*} = 90\,$m (with
$\rm \varepsilon_{n} = 3.75\,\mu m\,rad$) and $\beta^{*} = 1540\,$m (with
$\rm \varepsilon_{n} = 1\,\mu m\,rad$), respectively.
At the blue line, the Coulomb and the nuclear scattering amplitudes have equal 
moduli.
}
   \label{fig:coulomb}
\end{figure}

The comparison of the RP acceptances in $t$ for the two  
high-$\beta^{*}$ optics foreseen (Figure~\ref{fig:elasticaccept}) shows the
significantly better reach of the $\beta^{*} = 1540\,$m to low $|t|$ as
compared to 90\,m. In both cases, the lower centre-of-mass energy is 
advantageous: reducing $\sqrt{s}$ from 14\,TeV to 7\,TeV results in a factor
0.5 in the lowest reachable $|t|$-value (see Figure~\ref{fig:coulomb} 
and~\cite{lhcc2010}).

\subsection{Extrapolation at \MakeLowercase{$\beta^{*}=90\,$m}}
The contributions to the extrapolation uncertainty at $\sqrt{s} = 7\,$TeV are
expected to be very similar to those at $\sqrt{s} = 14\,$TeV discussed 
in~\cite{JINST}. 
The key 
advantage of the lower energy, however, is the about 50\% shorter extrapolation
interval, which will reduce the fit-induced statistical error and the 
systematic uncertainty contribution from the model dependence of the 
extrapolation function.
The beam divergence at 7\,TeV is higher than at 14\,TeV by a factor 
$\sqrt{14/7} = \sqrt{2}$ (i.e. $3.3\,\mu$rad
instead of $2.4\,\mu$rad), but as explained in~\cite{JINST} (Section~6.3.2), 
this effect leads to a shift in the extrapolation
result which can be corrected for.
The optical functions are expected to be known to within $\sim $1\%, which 
would translate into an extrapolation uncertainty contribution of 1.5\%.
The statistical uncertainty after an 8 hour long run -- with conditions 
explained in the next section and Table~\ref{tab:scenario90m} 
-- would be on the 1\% level.

In summary, a total error estimate of $\sim 2$\% for the extrapolation 
at $\beta^{*}=90$\,m and 7\,TeV can be considered as conservative.

\subsection{Running Scenario for the 
\MakeLowercase{$\beta^{*}=90\,$m} Optics in 2011}
After the commissioning of the unsqueeze from the injection optics to 
$\beta^{*}=90\,$m~\cite{burkhardt}, TOTEM is aiming at four well separated 
runs of typically 8\,hours length, enabling systematic comparisons between the
individual results and successive improvement of the beam conditions.
As outlined above, the key element in reducing the extrapolation uncertainty
is the minimisation of the $|t|$-interval to be bridged, by extending the 
acceptance to the lowest possible $|t|$ and hence by moving the RPs as close
as possible to the beam.
As shown in 
Table~\ref{tab:scenario90m}, this is accomplished by a twofold strategy:
\begin{itemize}
\item Reduce $d_{RP} / \sigma_{beam}$. In special runs in 2010, successful data 
taking took place as close as 7\,$\sigma$ from the beam, i.e. in the shadow of only
the primary collimators.
For machine protection reasons, this approach imposes a limit on the total 
beam current\footnote{The quantitative current limit for $\beta^{*}=90\,$m runs 
remains to be decided by the Machine Protection Panel and the collimation group.}. 
\item Reduce the transverse beam emittance and hence the width 
$\sigma_{beam}$. Thus for a given $d_{RP} / \sigma_{beam}$, the pots are moved to a
smaller absolute distance $d_{RP}$.
\end{itemize}
The bunch population $N$ is not only limited by the low-current requirement but
most importantly in view of keeping the inelastic pile-up fraction 
$\mu_{inelastic}$ below 10\% which is essential for an accurate measurement
of the inelastic rate. With one bunch of $(6 \div 7) \times 10^{10}$\,p/b,
each of the four runs requested can provide an elastic extrapolation with
the required precision $\sim 1$\%.

\begin{table}[h!]
    \setlength\tabcolsep{4pt}
    \caption{\it Tentative scenario for the beam conditions in four physics runs
with the $\beta^{*} = 90\,$m optics at 3.5\,TeV beam 
energy\protect\footnotemark[2].
$d_{R\!P}$ denotes the distance between the 
outer window surface of the pots and the centre of the beam. The additional distance 
of 0.5\,mm to the sensitive detector volume has been taken into account in the calculation of $|t_{50}|$, i.e. the $|t|$-value where the acceptance reaches 
50\%. $\mu_{inelastic}$ is the inelastic pile-up fraction. 
$\delta \sigma_{el} (t = 0)$
is the precision of the elastic cross-section extrapolation to $t = 0$.}
    \label{tab:scenario90m}
    \begin{tabular}{|l|c|c|c|c|}
        \hline
Run                & 1 & 2 & 3 & 4 \\
\hline
$\varepsilon_{n}$ [$\mu$m\,rad]    & 3  & 3  & 1  & 1  \\
$d_{R\!P} / \sigma_{beam}$           & 8  & 6  & 8  & 6  \\
$N / 10^{10}$                      & 7  & 7  & 6  & 6  \\
$L$ [$10^{27}\rm cm^{-2} s^{-1}$]  & $6.1$  & $6.1$ &  $13$   & $13$\\
$\mu_{inelastic}$                  & 0.04 & 0.04  & 0.08  & 0.08  \\
$|t_{50}|$ [$\rm GeV^{2}$]         & 0.016 & 0.0096 & 0.0060  & 0.0037 \\
$\int_0^{8h} L\,\d t$ [nb$^{-1}$]    & 0.2  & 0.2    & 0.3    & 0.3    \\
$\delta \sigma_{el} / \sigma_{el} (t = 0)$  
                                   & $\sim 1.5$\% & $\sim 1$\% & $< 1$\% & $< 1$\% \\
\hline
    \end{tabular}
\end{table}
\footnotetext[2]{This table
has been modified w.r.t. the 4.0\,TeV version presented in the workshop 
slides.}

\section{COMBINED UNCERTAINTY AT \MakeLowercase{$\mathbf{\beta^{*}=90\,\mathrm{m}}$}}
The error contributions from all measured quantities and from the theoretical 
knowledge of $\rho$ are listed in Table~\ref{tab:combinederror}. The combined 
uncertainties in $\sigma_{tot}$ and $L$ result from an error propagation 
calculation taking into account the correlations. Note that the precision in 
$L$ is slightly worse due to its squared dependence on $(N_{el} + N_{inel})$,
as compared to the linear dependence of $\sigma_{tot}$ (see 
Eqns.~(\ref{eqn_sigmatot}) and~(\ref{eqn_lumi})). 

\begin{table}[h!]
    \setlength\tabcolsep{4pt}
    \caption{\it Error contributions from all ingredients to Eqns.~(\ref{eqn_sigmatot}) and~(\ref{eqn_lumi}) for $\beta^{*}=90\,$m.}
    \label{tab:combinederror}
    \begin{tabular}{|l|l|}
        \hline
Extrapolation of $\frac{\d\sigma_{elastic}}{\d t}$ to $t = 0$ & $\pm 2$\% \\
Total elastic rate   &  \\
(strongly correlated with extrapolation)   & $\pm 1$\% \\
Total inelastic rate                                          & $\pm 1.4$\% \\
Error contribution from $(1+\rho^{2})$ using full & \\
COMPETE error band 
$\delta\rho/\rho = 33$\% & $\pm 1.2$\% \\
        \hline
Total uncertainty in $\sigma_{tot}$ incl. correlations & $\pm 3$\% \\
Total uncertainty in $L$ incl. correlations & $\pm 4$\% \\
        \hline
    \end{tabular}
\end{table}

\section{Outlook: Performance at \MakeLowercase{$\beta^{*}=1540\,$m}}
The $\beta^{*}=1540\,$m optics, studied in detail for 
$\sqrt{s} = 14\,$TeV~\cite{JINST}, extends the measurable $|t|$-range to 
values of the order $\rm 10^{-3}\,GeV^{2}$ and thus enables an elastic 
extrapolation with an uncertainty at the 0.2\% level. Consequently, the 
precision in $\sigma_{tot}$ improves to $\sim 1$\%.

Scaling the optics properties to $\sqrt{s} = 7\,$TeV shows an even further 
enhancement of the acceptance reach to low $|t|$ (Figure~\ref{fig:coulomb}, 
bottom block of curves). Inserting the RPs to 8 or 6
$\sigma_{beam}$ would give access to the elastic scattering zone 
dominated by the
Coulomb interaction and thus permit a measurement of $\rho$, avoiding any need
for theoretical input to the $\sigma_{tot}$ determination via the Optical 
Theorem. At $\sqrt{s} = 14\,$TeV this opportunity will not be offered.

However, the current version of the $\beta^{*} = 1540\,$m optics for TOTEM is 
only compatible with operation at $\sqrt{s}=10\,$TeV to 
14\,TeV~\cite{burkhardt}.
At lower energy the two main limiting parameters are the minimum strength 
allowed in the insertion quadrupoles
and the aperture. One way to avoid these limitations would be to loosen the 
constraints on the phase advance by
abandoning the condition to have ``parallel-to-point focussing'' in both 
transverse projections.
Alternative optics at very high $\beta^{*}$, compatible 
with operations at $\sqrt{s}=7$\,TeV are under study.

\end{document}